# Metallophilicity Enhances Electron Transport through Parallel Organometallic 1D Chain Junctions Formed In Situ

Sigifredo Luna Jr.,[a] Hannah E Skipper,[a] Brent Lawson,[b] Eric Cueny,[a] and Maria Kamenetska[a,b,c*]

[a] Department of Chemistry, Boston University, Boston, Massachusetts, 02215, United States

[b] Department of Physics, Boston University, Boston, Massachusetts, 02215, United States

[c] Division of Material Science and Engineering, Boston University, Boston, Massachusetts, 02215, United States

E-mail: mkamenet@bu.edu

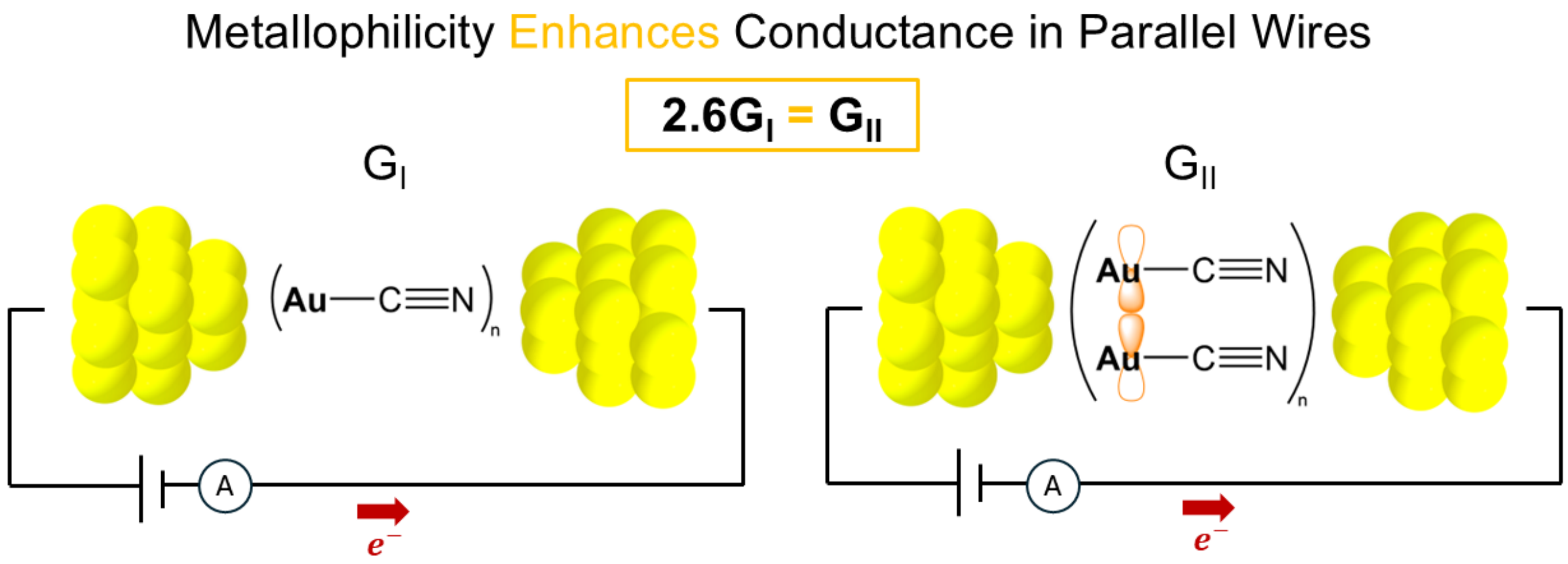

**Abstract**

We reveal the role of aurophilic interactions in the formation and conductance of gold cyanide molecular wires of variable length-to-width ratios assembled at the tip of an STM break junction in ambient conditions. Specifically, we identify electron transport signatures through 1D single chains containing variable number of monomeric repeats of gold cyanide AuCN, linked in series $(AuCN)_n$, and through adjacent molecular wires linked in parallel. When bound in series, destructive quantum interference causes an exponential decay of conductance in $(AuCN)_n$ 1D wires for n=1-3. But when bound in parallel, aurophilic coupling through the gold atoms of neighboring chains reorders electronic states and results in significant enhancement of conductance. Our work reveals that metallophilicity can play a significant role in junction assembly and electron transport characteristics.

## Introduction

Identifying and finding synthetic routes for novel reduced-dimension materials is of interest for emerging technologies such as nanoscale electronics and quantum materials.[1,2] Currently, top-down methods, such as lithography, for generating nanoscale morphologies predominate but are limited to feature sizes of ~10 nm or greater. While 2D materials can be made atomically flat using exfoliation methods, no analogous techniques exist for 1D material preparation in room temperature conditions.[3] Yet, atomic or molecule 1D structures containing both metal and organic components, known as 1D or quasi-1D coordination chains respectively, are particularly appealing for electronic and spintronic applications.[4–7] Synthetic techniques can be used to assemble such chains in ambient conditions but usually only when packed into 3D crystals, making it challenging to study electron transport in intact extended 1D chains and to disentangle the effects of inter-chain coupling.[8–10]

Bottom-up approaches for forming 1D or quasi-1D structures are an alternative to bulk synthesis or lithography. Recent cryogenic studies using scanning tunneling microscopy (STM) show that isolated quasi-1D coordination chains can assemble from organic precursors on an atomically flat metal substrate in vacuum by incorporating, through oxidation, gold atoms from the surface.[11] In contrast, other coordination complexes are susceptible to molecular fragmentation on gold surfaces.[12–14] Our prior work using Scanning Tunneling Microscope break junction (STMBJ) approaches in ambient conditions shows that certain coordination complexes are labile in the STM break junction[15] and that some assembly of other short chains can occur.[16–18] Here, we add aurophilicity[19,20] —an attractive interaction between gold metal ions—into the chemical toolkit of molecular electronics to promote formation of a new class of molecular circuits containing repeat units of cyanometallates of distinct length-to-width ratios

Cyanometallates are a class of anionic transition metal complexes coordinated by cyanide ligands ($CN^-$) through covalent M-CN bonding. The complex $[Au(CN)_2]^-$ shown in Figure 1A (top) is a linear unit featuring a two-coordinate Au(I) center. The Au(I) forms σ bonds through the Au *s* or $dz^2$ orbitals and π back-bonds from $d^{10}$ Au(I) to the CαN π* as shown in Figure 1B (top). Prior measurements reveal this bond to be highly robust, ~ 4 eV by some estimates,[21] which is comparable to C-C bonds, suggesting that the complex should withstand the forces inherent in junctions and on rough metal surfaces unlike other complexes.[15,22] In addition, the 6p orbitals on the Au(I) metal center are part of the LUMO manifold, but are known to become partially filled under some conditions when two adjacent Au(I) centers form an Au(I)-Au(I) bond, an interaction known as aurophilicity, shown in Figure 1B (bottom).[19,20] The N lone pair on the terminal nitrogen of the cyano group provides another source of intermolecular donor-acceptor interactions. Overall, cyanometallates, including $[Au(CN)_2]^-$, are common units in the formation of extended 3D structures as shown in Figure 1C.[23,24] Separately, cyanide (-C≡N)[25,26] and isocyanide (-N≡C)[27] functional groups installed on short organic backbones have been previously reported to bind to Au electrodes and promote molecular junction formation.

Here, we demonstrate step-wise assembly of single covalently bound cyanometallate molecular wires of variable length at the tip of an STM, in situ, in ambient conditions using the STMBJ approach. Our results indicate that aurophilicity promotes assembly of wires in parallel and boosts conductance. We characterize electron transport through single and multiple adjacent wires containing variable number (n=1-3) of monomeric repeats of gold cyanide (AuCN) and quantify the changes in transport resulting from the addition of material monomers along two distinct directions, along and perpendicular to the direction of transport. Through a combination of experimental measurements, density functional theory (DFT), and non-equilibrium Green's function (NEGF) transmission calculations, we determine that the conductance of chains of $(AuCN)_n$ decays with chain length due to destructive interference. In addition, adjacent wires couple through the gold metal centers to form Au(I)-Au(I) inter-chain bonds, which is evidence for the first time the role of aurophilicity in junction formation. Importantly, these inter-chain aurophilic interactions result in enhanced electron transport through the parallel molecular wires, yielding more than integer multiples of the monomeric material conductance. These results point to new possibilities for forming reduced dimension materials using unique properties of Au-containing complexes and for tuning transport and quantum interference using metallophilic bonds.

## Results and Discussion

Single molecule conductance measurements are performed using a home-built STMBJ as previously described and briefed in the methods.[18,28] We perform single molecule conductance measurements by drop casting potassium dicyanoaurate(I) $K[Au(CN)_2]$ on a gold substrate out of 1 mM aqueous solutions.[18] The resulting one-dimensional (1D) log and linear-binned conductance histograms shown in Figure 2A and Figure S1, respectively, reveal that the presence of $K[Au(CN)_2]$ on Au electrodes results in numerous conductance features ranging from $\sim 10^{-1} – 10^{-6}$ $G_0$. The signatures are insensitive to bias voltage below 500 mV as we show in the SI Figure S2.

Based on prior work[18,29] and these results, we conclude that the salt dissociates into constituent ions, so that the $[Au(CN)_2]^-$ can bind to gold electrodes. Metal-molecule junctions formed through donor-acceptor amine and imine-gold binding are well established but usually feature a single or, at most, two conductance signatures.[1,30,31] Here, in the presence of $[Au(CN)_2]^-$, we identify *eight* distinct peaks, indicated by arrows in Figure 2A. We group them into three conductance regimes of relative "high" conductance (HG), "medium" conductance (MG) and "low" conductance (LG) and label the distinct conductance features from lowest to highest within each regime. The characteristic conductance of each feature, listed in Table 1, is determined by fitting multiple Gaussian curves to the linear-binned histogram of each region (Figure S1).

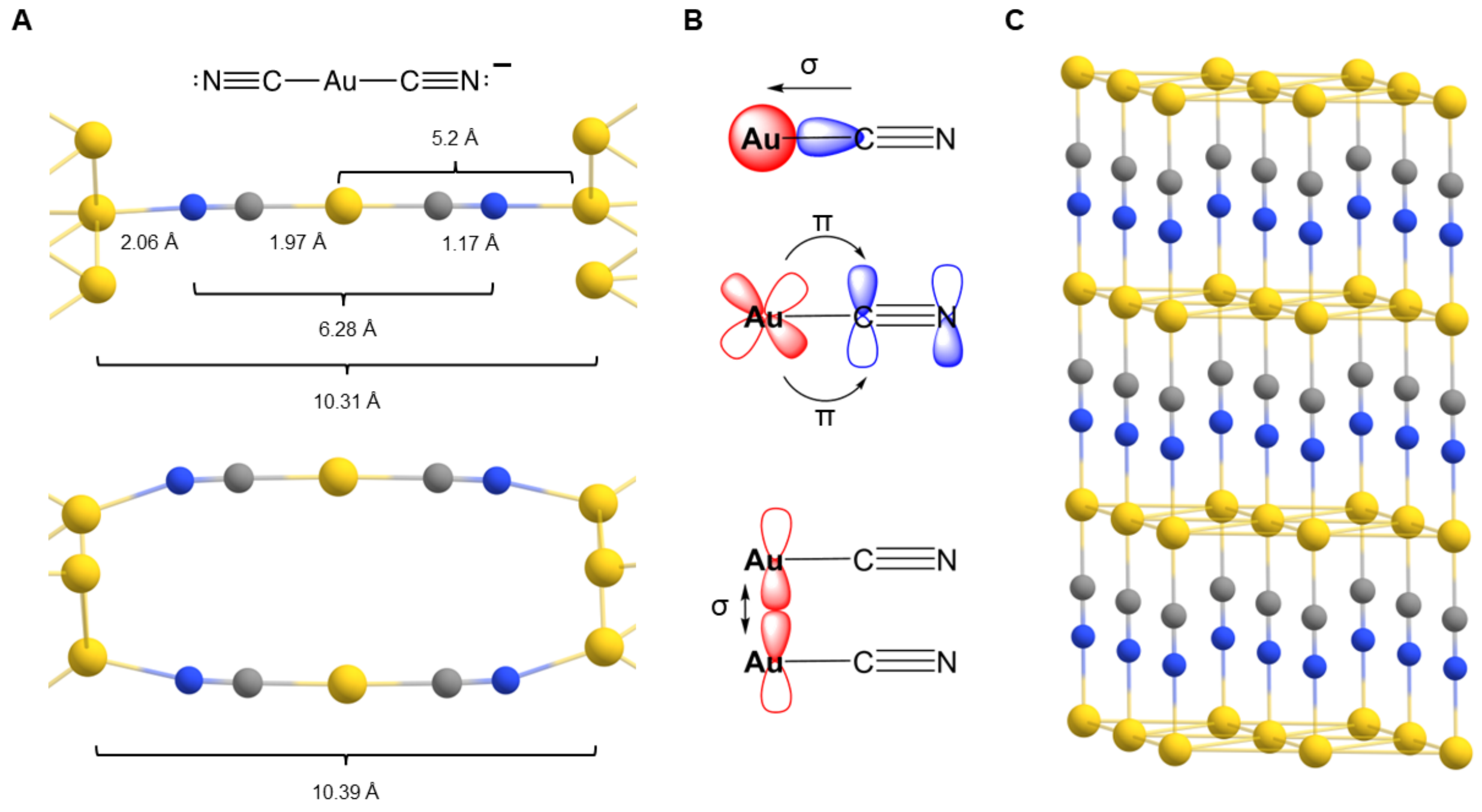

**Figure 1.** A) The structure of the $[Au(CN)_2]^-$ ion (top) and corresponding lengths as predicted by DFT when bound in the junction as monomer (top) and as dimer (bottom) . B) Schemes of intra (top) and intermolecular (bottom) $Au(CN)_2$ interactions; the Au-CN bond is formed through σ bonding involving Au 6s-5$d_{z^2}$ orbital as well as through π back-donation to the Au *6d* which reinforces the interaction; complexes can also interact through the Au(I) center *6p* orbitals which can become partially filled to form aurophilic *6pσ* bonds. C) 3D crystal of aurocyanide based on neutron diffraction ref [23].

Two-dimensional (2D) histograms display conductance as a functional of electrode displacement, which provide simultaneous length and conductance information.[28,32] From the 2D histogram in Figure 2B it is apparent that the junction lengths of these three regions are distinct. We identify characteristic elongation lengths for the HG, MG and LG regions as marked in dashed vertical lines and list them in Table 1. We consider whether a single $[Au(CN)_2]^-$ anion can generate all of the features observed. The HG region has the highest conductance and shortest junction length with plateaus extending to ~ 0.2 nm, while the LG region has the lowest conductance and longest junction length with plateaus extending to ~ 1.1 nm. Accounting for the Au snapback distance of ~ 0.5 – 0.8 nm,[28,33] these features correspond to molecular junction lengths of ~ 0.7– 1 nm and ~ 1.6 – 1.9 nm respectively. Importantly, the lengths of the LG junctions are significantly longer than the length of the entire complex as shown in Figure 1A. Therefore, we conclude that a single $[Au(CN)_2]^-$ anion alone cannot generate features in the LG regime without rearrangement or appending of Au atoms from the electrodes into the molecular junction bridge.

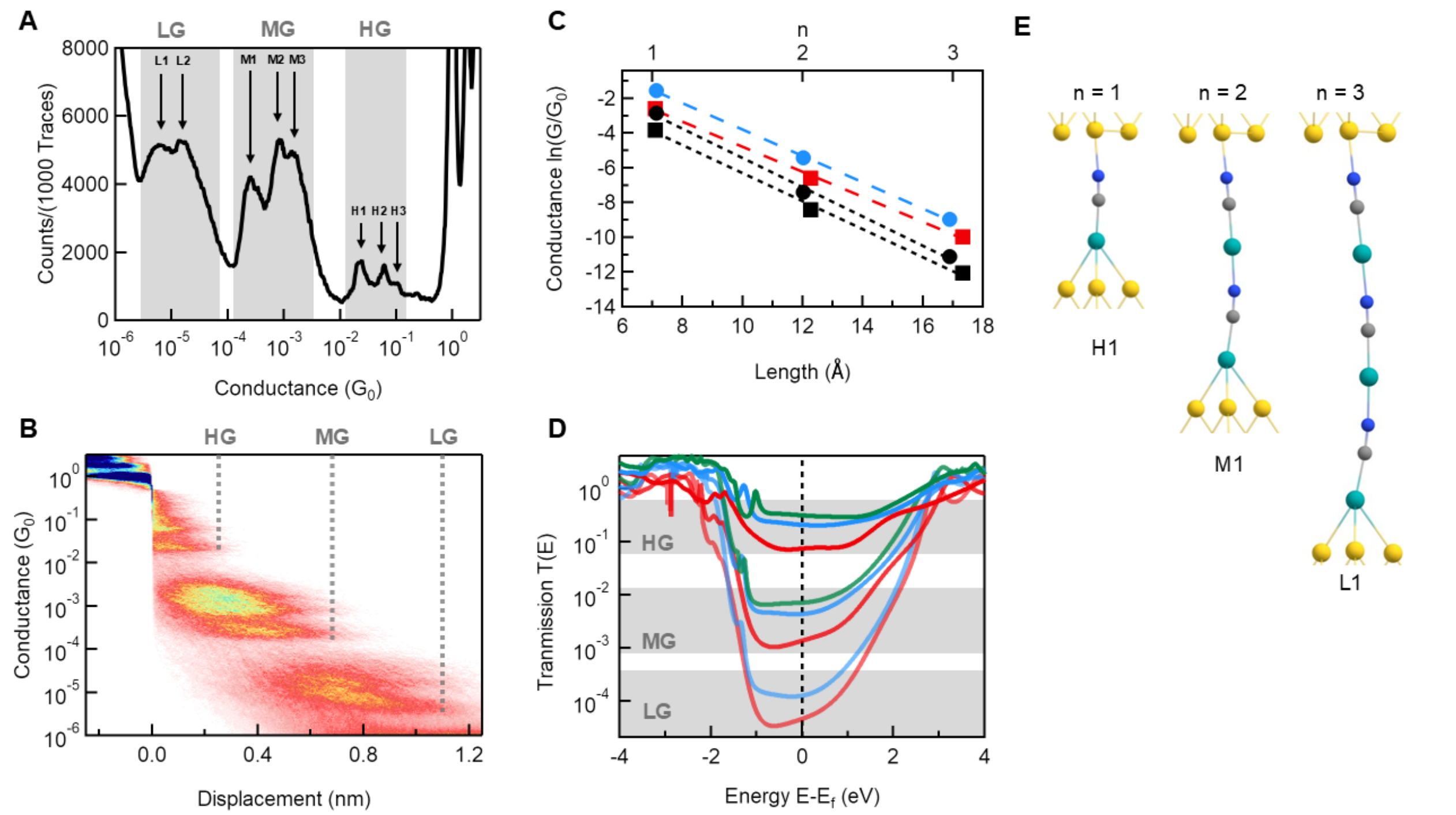

**Figure 2.** (A) 1-D log-binned and (B) 2-D conductance histograms from measurements in the presence of 1 mM $[KAu(CN)_2]$ at 500 mV. (C) Conductance as measured for peaks H1, M1 and L1 (black squares); H2, M2 and L2 (black circles) and DFT-calculated transmission of $\{Au(CN)\}_n$ for *n=1-3* monomers (red square) and dimers (blue circles) versus the number of repeating units *n* and the length, in Angstroms, of the junction as calculated in DFT from edge to edge of the apex electrode Au atoms. (D) DFT calculated transmission curves and (E) relaxed structures for $\{Au(CN)\}_n$ *n = 1-3* with $Au_{32}$ electrodes (yellow) and Au(I) gold atoms (teal). Transmission curves corresponding to the structures H1, M1 and L1 shown are plotted in red.

The distance between the dashed lines in Figure 2B is ~ 4.5 Å, indicating that the characteristic length of each distinct region increases in multiples of this unit length. Therefore, we hypothesize that a discrete molecular unit ~4.5 Å is repeatedly appended into the junction during elongation, increasing the length of the molecular chain spanning the Au-Au gap.[27] We consider the lengths of the molecular units present in the junction which can account of this experimental observation. As seen in Figure 1A, the neutral unit (AuCN) measures ~ 3.2 Å, and ~ 5.2 Å if we include the bond to the apex Au atom of the electrode, which is a more reasonable match for the extension increase of ~4.5 Å we find above. These results suggest that discrete units of the molecular motif (AuCN) are appended in series to create 1D chains in the junction, resulting in sequential drops of conductance during junction elongation from the HG, to MG and finally to the LG.

The established transport mechanism through close-shell molecular junctions is non-resonant tunneling, where conductance $G$ is known to decay exponentially at a rate $\beta$ as molecular length is increased in units *of* $L_0$:[34,35]

$$G \propto e^{-\beta n L_0}$$

Here $n$ is an integer. We examine the conductance decay observed here. We follow literature precedent to interpret the most extended conductance signatures H1, M1 and L1 within each region as the geometry which corresponds to a single molecular bridge fully stretched between electrodes and with no additional molecules bound in parallel.[27] In Figure 2C, the conductance values of H1,

M1, and L1 are plotted in black squares against the number of repeating units assembled in series, *n* (top x axis), and display exponential decay, in agreement with the non-resonant transport model above.

**Table 1.** Experimental and calculated conductance and displacement information

| | | Experiment | | Theory | |
|---|---|---|---|---|---|
| Region | Peak | Conductance ($G_0$) | Junction lengths (nm) [a] | Transmission | Au-Au Extension (nm) [b] |
| HG | H3 | $9.1 \times 10^{-2}$ | | $3.2 \times 10^{-1}$ | 0.72 |
| | H2 | $5.7 \times 10^{-2}$ | | $2.1 \times 10^{-1}$ | 0.71 |
| | H1 | $2.2 \times 10^{-2}$ | 0.7 | $7.5 \times 10^{-2}$ | 0.71 |
| MG | M3 | $1.4 \times 10^{-3}$ | | $7.2 \times 10^{-3}$ | 1.21 |
| | M2 | $6.0 \times 10^{-4}$ | | $4.4 \times 10^{-3}$ | 1.20 |
| | M1 | $2.2 \times 10^{-4}$ | 1.1 | $1.4 \times 10^{-3}$ | 1.23 |
| LG | L2 | $1.5 \times 10^{-5}$ | | $1.3 \times 10^{-4}$ | 1.69 |
| | L1 | $5.8 \times 10^{-6}$ | 1.6 | $4.7 \times 10^{-5}$ | 1.73 |

[a] Exp plateau lengths include snapback of ~ 0.5 nm

[b] Measured edge to edge between Au electrode trimer tips

We use DFT and NEGF transmission calculations to explore possible junction geometries. The resulting transmission spectra are plotted in Figure 2D. We first consider monomer junctions of $Au_{32}$-$(AuCN)_n$-$Au_{32}$ where $n$ = 1, 2, or 3 as shown in Figure 2E. The molecular wires $(AuCN)_n$ are modeled with Au(I) atom(s) (teal). We note that the Au atom originally part of the (AuCN) unit forms an apex tip atom bound to a blunt trimer tip. Analogous junctions formed on sharp electrodes were also constructed (Figure S3 in the SI) and yielded qualitatively identical results. We find that all three proposed molecular wires with $n$ = 1-3 form stable junctions with the energy required to break a N-Au bond equal to 1.1, 1.5 and 1.6 eV, respectively. These calculated binding energies to the Au electrodes are significantly larger than typical donor acceptor binding energies in molecular junctions such as amine (-$HN_2$, 0.6 eV), thioether (-SMe, 0.5 eV) and pyridyl (-py, 1 eV) groups,[36–38] and are more consistent with reports of charged species self-assembling in the junction, such as imidazolate[18] and 1,4-phenylene diisocyanide, which also assemble into coordination chains in the junction.[27] These binding energies are also comparable to the inter-atom Au-Au binding within the electrodes which has previously been measured to be ~1.5 eV.[39] We conclude that the $CN^-$-Au binding is sufficiently robust to pull Au atoms out of the electrodes and into the junction during stretching. From the minimum energy DFT geometries shown in Figure 2E, we determine the distance between the blunt $Au_{32}$ electrodes to be 7.1, 12.3, and 17.3 Å, for $n$ = 1-3 respectively and report them in Table 1. The difference in extension between these geometries is ~ 5 Å, which is in good agreement to the experimentally determined values of ~ 4.5 Å.

From the transmission curves in Figure 2D, we extract the predicted low-bias conductance of geometries in 2E as the transmission at Fermi ($E-E_f = 0$) and plot it in Figure 2C in red squares alongside the experimentally measured conductance in black squares. We note that the calculated transmission at $E_f$ is ~ 50% higher than the experimentally determined conductance for H1, M1, and L1, consistent with the widely established tendency of DFT to over-estimate conductance of metal-molecule junctions.[30,40] Importantly, while the absolute value of the DFT-calculated transmission is overestimated, the trends between junction geometries are known to be representative and reliable. Here, we find a DFT-predicted conductance decay for the $(AuCN)_n$ wires to be $\beta = 0.72$ Å$^{-1}$, in excellent agreement with the experimentally determined molecular $\beta = 0.80$ Å$^{-1}$ which we obtain from the fit to the data in Figure 2C.

This agreement between the calculated conductance of a series of $(AuCN)_n$ wires and our measurements supports our interpretation that the deposited $[Au(CN)_2]^-$ rearranges in the junction and assembles into 1D wires of various lengths at the tip of the STM during elongations. The original $[Au(CN)_2]^-$ anion can also bind without rearrangement, as shown in Figure S3, and has a similar conductance to the n=2 structure. Following literature precedent, the secondary and tertiary peaks in each region, H2 and M2 for example, correspond to two or three 1D chains of different lengths bound *in parallel.* Such features have been observed for imidazole and isocyanide linkers and displayed a conductance at integer multiples of the primary peak.[18,27] Strikingly, here the secondary peaks in each region H2, M2 and L2, display *greater* than integer conductance of H1, M1 and L1, respectively. The multiple, listed in Table 2, is ~2.6 in each case, resulting in the same $\beta$ decay constant observed for the dimer peaks in experiment (black circles) and in theoretical calculations (blue circles) as shown in Figure 2C. The similarity in β between the primary and secondary peaks suggests that the same 1D material is responsible for both conductance features. We conclude that the addition of monomer units *in parallel* to make dimers and trimers in the junction, causes inter-chain coupling and electronic rearrangement which boosts conductance relative to the monomer.

To investigate the cause of these conductance trends, we study the electronic structure of various wires within junctions. We examine the frontier molecular orbitals of the gas phase $[Au(CN)_2]^-$ anion, shown in Figure 3B and Figure S4-S5. We find that the molecular HOMO is composed of σ orbital overlap between the Au $6s$-$5d_{z^2}$ atomic orbitals and the cyano σ as expected.[41,42] In contrast, according to the natural bond order (NBO) analysis[43–45] (Figure S6), the LUMO is dominated by the Au *6p*. Eigenstate analysis of the monomer junction in Figure 3A, (Figure S7) reveals that the junction transmission at $E_f$ is dominated by the state E1, which has the symmetry of the gas-phase molecular HOMO. Similarly, the Au *6p* and *d*-derived states, such as E2 and E3 respectively in Figure 3C and S, are part of the LUMO manifold. It is interesting to note that the symmetry of these frontier MO's which dominate transport implies destructive quantum interference (DQI) through the molecule,[46] which is observed as a pronounced dip in the transmission spectrum at Fermi and can account for the high decay rate $\beta$ of ~0.80 $Å^{-1}$ we found above. Surprisingly, this decay is similar to what is observed in saturated alkane wires.[47–53] We conclude that, despite containing transition metal ions Au(I), the AuCN wires are insulating. On the other hand, the presence of Au seems to boost conductance in parallel wire junctions.

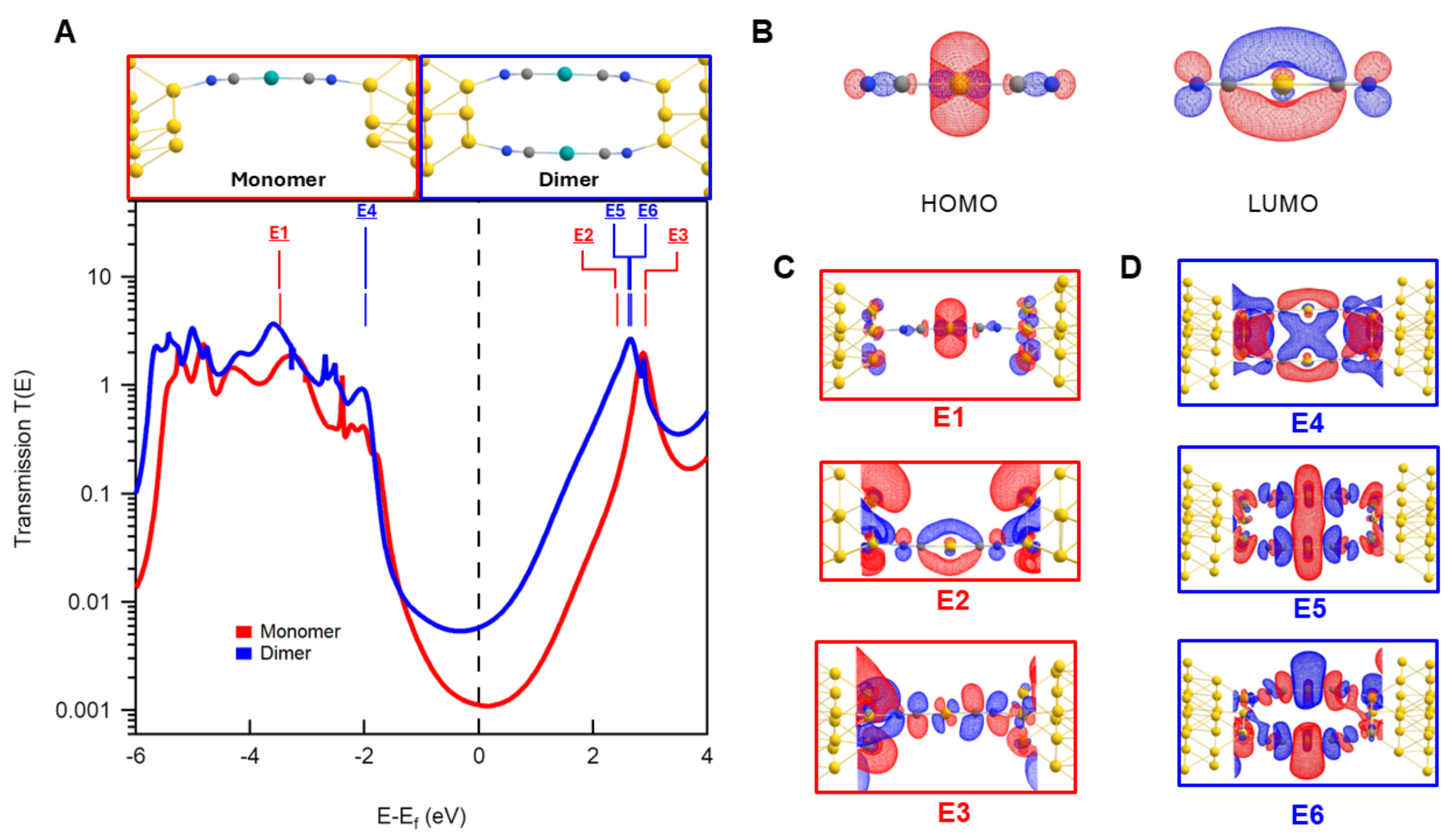

Figure 3: A) (Top) DFT relaxed structures and (Bottom) Transmission spectra for Monomer and Dimer $[Au(CN)_2]^-$ wires between $Au_{32}$ electrodes B) Isosurface plots of frontier MOs of $[Au(CN)_2]^-$ at 0.1 contour level C) Isosurface plots of select eigenstates of Monomer structure (see Figure S7 for contour levels) D) Isosurface plots of select eigenstates of Blunt Dimer structure (see Figure S8 for contour levels).

Now switching to the parallel dimer in Figure 3A bottom, we observe that its two constituent monomers couple to each other electronically through the Au(I) centers. In particular, the E4 HOMO eigenstate of the dimer is a symmetric combination of the E2 monomer states, with a σ bond between the two Au(I) centers. An NBO analysis in Figure S6 reveals that both the E2 and E4 states in the $LUMO_{monomer}$ and $HOMO_{dimer}$ manifolds respectively are composed almost exclusively of π symmetry, with a 52% and 47% contribution of the Au 6p states. Such a 6*p* σ overlap between the Au(I) atoms and its filled character as part of HOMO, is a signature of aurophilic interactions in binuclear complexes and dimeric structures.[54] In contrast, the σ Au 6s-$5d_{z^2}$ derived MO's, E5 and E6, become part of the LUMO. The frontier orbitals E4 and E5 contribute most to transport and reduce DQI features in the transmission spectrum of the dimer compared to the monomer. We conclude that the unexpected conductance enhancement observed here in dimer junctions is due to inter-molecular Au-Au σ bonding, known as aurophilic interaction, which shifts the 6p states from the LUMO to the HOMO manifold and further rearranges the spectrum, mitigating DQI and resulting in more than double the conductance at the Fermi energy. (See Figure S8 discussion for transmission comparison for junctions in figure 3A).

**Table 2**. Conductance ratios between distinct conductance features as measured (Figure 2A) and calculated at Fermi (Figure 2D).

| Feature Comparison | Conductance Ratio | Transmission Ratio |
|---|---|---|
| H3/H1 | 4.1 | 4.2 |
| H2/H1 | 2.6 | 2.8 |
| M3/M1 | 6.4 | 5.2 |
| M2/M1 | 2.7 | 3.2 |
| L2/L1 | 2.6 | 2.7 |

Junctions containing a single (AuCN) units with conductance in the HG region cannot couple through central Au(I) atoms as shown in Figure 3. Instead, we bridge repeating units of $(AuCN)_n$ for n= 1, 2, 3, to electrodes and find common stable geometries of parallel wires as shown in Figure 4A where the *edge* Au(I) atoms, common to all regions H, M and L, couple. These atoms are shown in dashed lines. Similar results are obtained with Au(0) as shown in Figure S12. These atoms approach to within 2.92 angstroms, comparable to the known aurophilic radius.[55] The calculated transmission spectra are shown in Figure 4B. Significantly, for *all* regions, the calculated transmission and enhancement ratios for both dimers and timers are in close quantitative agreement to the measured values, as reported in Table 2. In all junctions, the LUMO has an outsized contribution at $E_f$, while the HOMO resonances fall off sharply and display a DQI feature at ~-0.5eV. Zooming into the frontier LUMO resonances that dominate transport in our dimer junctions, H2, M2 and L2 in Figure 4C top, middle and bottom respectively, we observe significant electron density in the inter-molecular space between the *edge* Au atoms (long the dashed lines).

See SI Figures S9-S14 for a comprehensive eigenstate analysis. Similar features are observed in the trimer junctions. Comparing these resonances of the dimer junctions to the gas phase Au 6p-derived MO's (Figure S6), we conclude that this density originates primarily from the Au(I) 6p-derived orbital, with some contribution from higher energy states, resulting in a σ bond between two Au(I) atoms in different chains. Importantly, this type of internuclear Au(I)-Au(I) 6pσ overlap, characteristic of aurophilic interactions is universal across all parallel junctions considered in Figure 4 and can account for the conductance enhancement in H, M and L regions observed experimentally. We conclude that aurophilicity determines transport trends and results in enhancement of parallel molecular chains, revealing its role in electron transport in organometallic materials and suggesting at the potential to engineer molecular junction transport phenomena using metallophilic interactions.

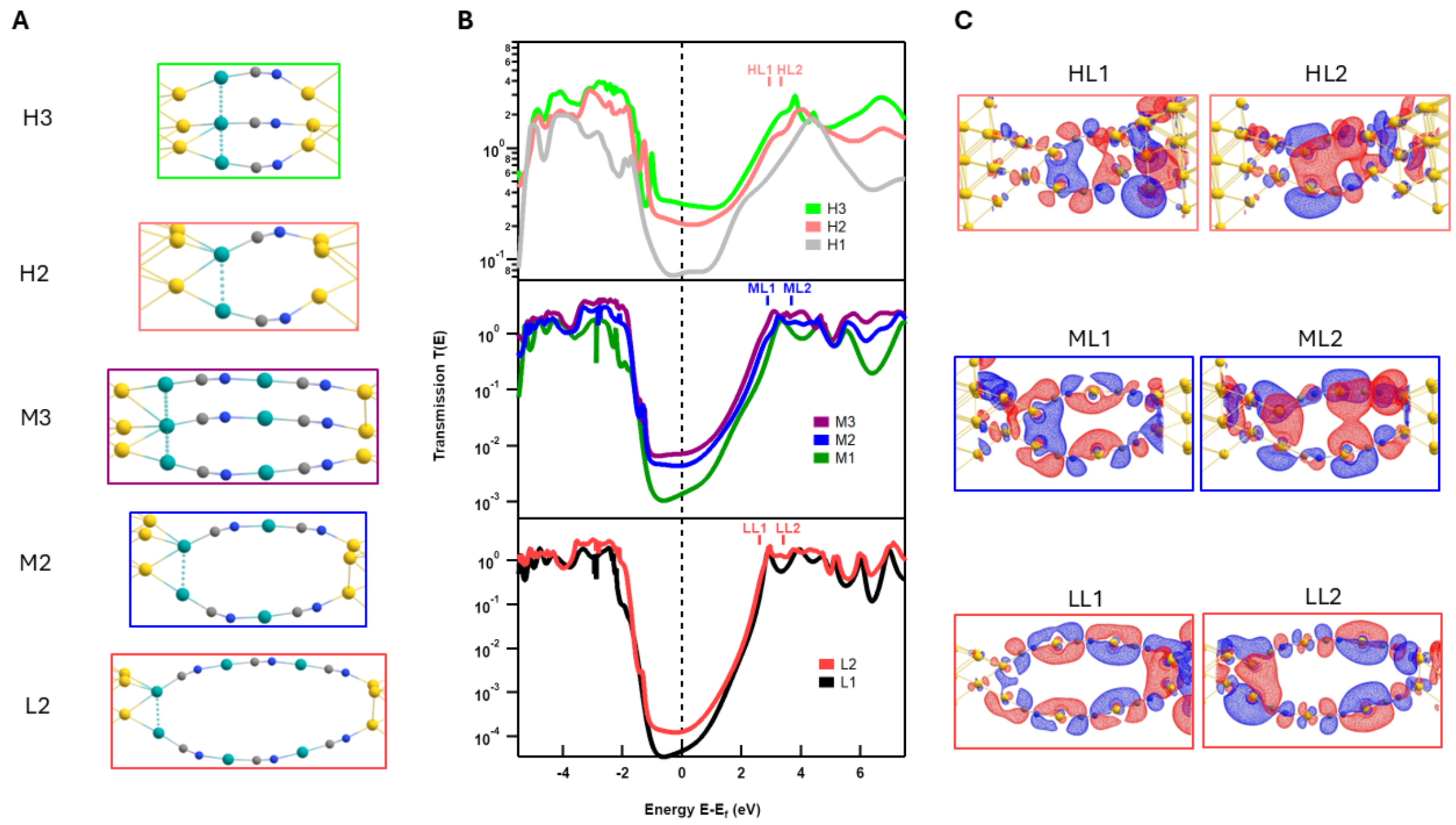

**Figure 4**: A) Proposed DFT relaxed structures of polymer wires of $(AuCN)_n$ in parallel with $Au_{32}$ electrodes. B) Transmission curves of HG, MG, and LG structures in the top, middle, and bottom plots, respectively. C) Selected Eigenstates of H2 (HL1, HL2), M2 (ML1, ML2), and L3 (LL1, LL2) with contour levels of 0.048, 0.028, and 0.028, respectively.

## Conclusions

We find that AuCN units form polymeric $(AuCN)_n$ complexes of different lengths that can assemble and bind in parallel in the junction. The edge Au atoms of these parallel wires form aurophilic interactions through the Au(I) 6p atomic orbitals and result in electronic rearrangement and significant conductance enhancement. We find that the single wires are insulating due to DQI, but the interference is offset by the aurophilic coupling which rearranges the electronic structure of parallel molecular bridges. Our results establish electronic signatures of aurophilicity in a single molecule junction and reveal how inter and intramolecular interactions in coordination complexes can be tuned to produce variable 1D structures containing transition metal atoms at the tip of an STM in ambient conditions. This study opens the door to leveraging metallophilicity to enhance transport and build functionality in molecular devises.

## Methods

### Experimental Details

Briefly, we bring an Au tip electrode in and out of contact with an Au substrate under a constant bias of 500 mV and record the conductance (current/voltage) as a function of the relative displacement of electrodes. Conductance traces recorded during junction stretching show plateaus at integer multiples of $G_0$ ($2e^2/h$) which correspond to the formation of Au point contacts with the gradually decreased number of Au atoms in the junction cross-section. After the Au contact is broken, a nano-gap is generated between electrodes and a molecule can bridge the gap to form a molecular junction. Aqueous solution of $K[Au(CN)_2]$ (~1mM) is deposited on the sample and dried under gentle heating of ~45˚C. We record at least 6000 conductance-displacement traces at constant bias voltages on the prepared sample and compile them into log-binned conductance histograms without data selection.

Computational methods

DFT calculations, unless otherwise stated, are performed with the FHI-aims software package[56] using the Perdew-Bruke-Ernzerhof (PBE) as the exchange-correlation functional[57]. Gas-phase and junction geometry relaxation calculations were performed with the light (similar to double-ζ) for all atoms for the $(AuCN)_n$ wires. Similarly, calculations for the $[Au(CN)_2]^-$ wires were performed with the pseudo-tight basis set. Candidate molecular junction geometries are constructed by bridging molecule(s) between Au[111] electrodes containing 32 Au atoms arranged in 4 layers with 3 apex atoms forming a trimer tip.[58] Each geometry is relaxed by allowing the molecule(s) and trimer Au atoms to rearrange while the 3 back layers for each electrode are frozen. To determine the most preferable inter-electrode distance, the electrodes are systematically moved in and out in steps of 0.05 Å, and the total energy is calculated. The junction with the lowest total energy is taken as a representative geometry for a given molecular bridge configuration. NEGF is used to calculate the energy-dependent transmission across the modeled junctions using the AITRANSS post-processor implemented within FHI-aims.[59–63]

The NBO analysis was performed using single-point gas-phase calculation within Gaussian[44] using the PBE functional[57] and the def2-QZVPD basis set[64–67]. The geometry of the molecular wires previously optimized within the junction in FHI-aims was directly used in the calculation. NBO Version 3.1, part of Gaussian software, was employed to perform the NBO analysis.[45] The Multiwfn software was used to obtain the natural atomic orbital (NAO) contribution.[43]

## Acknowledgements

This work was supported by the National Science Foundation under award #2145276.

**Supporting Information**

# Metallophilicity enhances electron transport through parallel organomatallic 1D chains formed in-situ

Sigifredo Luna Jr., Hannah E Skipper, Brent Lawson, Eric Cueny, Maria Kamenetska*

**Table of Contents**

## Methods

### *Experimental Details*

Briefly, we bring an Au tip electrode in and out of contact with an Au substrate under a constant bias of 500 mV and record the conductance (current/voltage) as a function of the relative displacement of electrodes. Conductance traces recorded during junction stretching show plateaus at integer multiples of $G_0$ ($2e^2/h$) which correspond to the formation of Au point contacts with the gradually decreased number of Au atoms in the junction cross-section. After the Au contact is broken, a nano-gap is generated between electrodes and a molecule can bridge the gap to form a molecular junction. Aqueous solution of $K[Au(CN)_2]$ (~1mM) is deposited on the sample and dried under gentle heating of ~45˚C. We record at least 6000 conductance-displacement traces at constant bias voltages on the prepared sample and compile them into log-binned conductance histograms without data selection.

### *Computational methods*

DFT calculations, unless otherwise stated, are performed with the FHI-aims software package[1] using the Perdew-Bruke-Ernzerhof (PBE) as the exchange-correlation functional[2]. Gas-phase and junction geometry relaxation calculations were performed with the light (similar to double-ζ) for all atoms for the $(AuCN)_n$ wires. Similarly, calculations for the $[Au(CN)_2]^-$ wires were performed with the pseudo-tight basis set. Candidate molecular junction geometries are constructed by bridging molecule(s) between Au[111] electrodes containing 32 Au atoms arranged in 4 layers with 3 apex atoms forming a trimer tip.[3] Each geometry is relaxed by allowing the molecule(s) and trimer Au atoms to rearrange while the 3 back layers for each electrode are frozen. To determine the most preferable inter-electrode distance, the electrodes are systematically moved in and out in steps of 0.05 Å, and the total energy is calculated. The junction with the lowest total energy is taken as a representative geometry for a given molecular bridge configuration. NEGF is used to calculate the energy-dependent transmission across the modeled junctions using the AITRANSS post-processor implemented within FHI-aims.[4–8]

The NBO analysis was performed using single-point gas-phase calculation within Gaussian[9] using the PBE functional[2] and the def2-QZVPD basis set[10–13]. The geometry of the molecular wires previously optimized within the junction in FHI-aims was directly used in the calculation. NBO Version 3.1, part of Gaussian software, was employed to perform the NBO analysis.[14] The Multiwfn software was used to obtain the natural atomic orbital (NAO) contribution.[15]

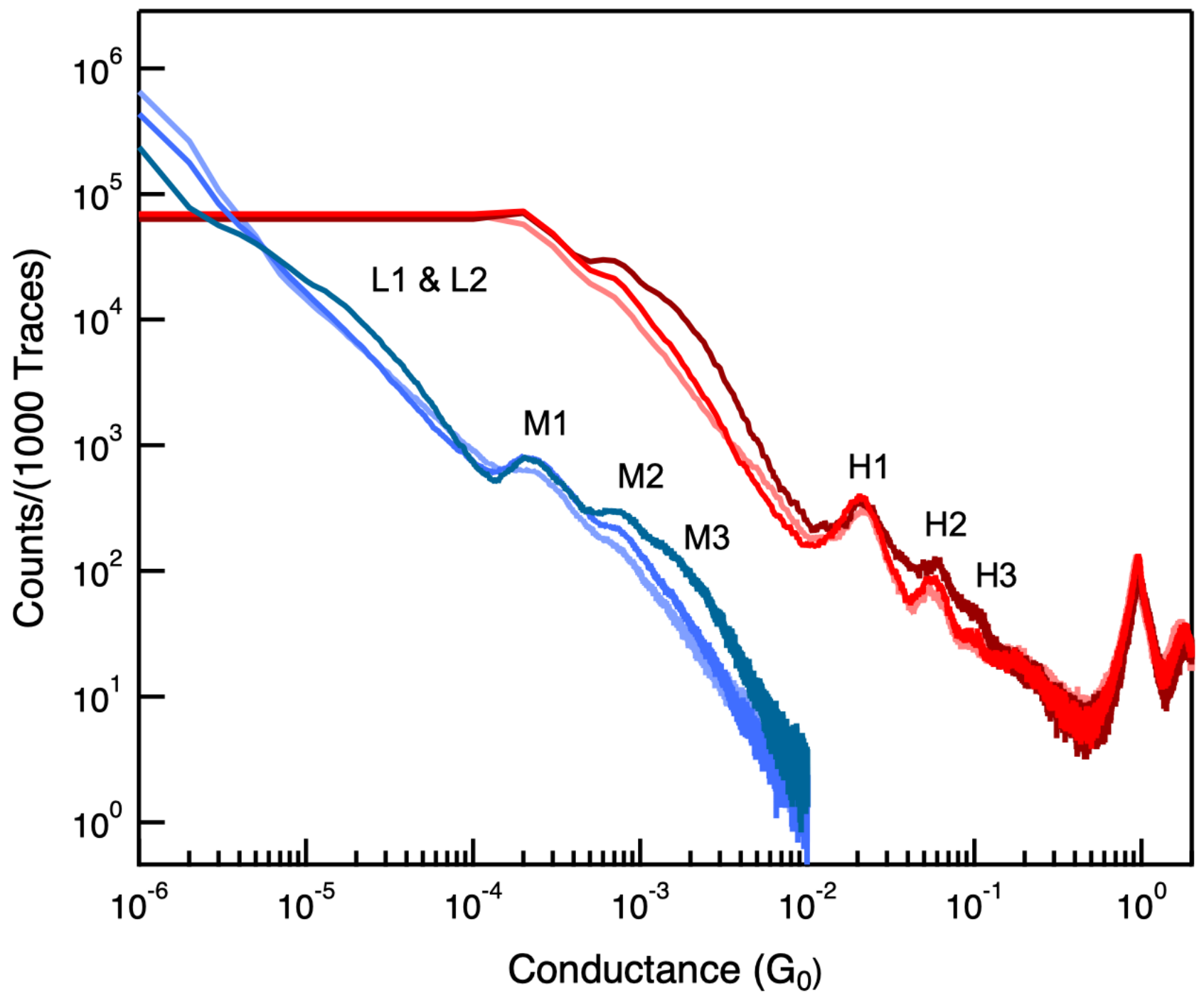

**Figure S1.** 1D linear-binned conductance histogram using bins of 1 x $10^{-4}$ (red) and 1 x $10^{-6}$ (blue). Different shades of red and blue traces indicate multiple data sets.

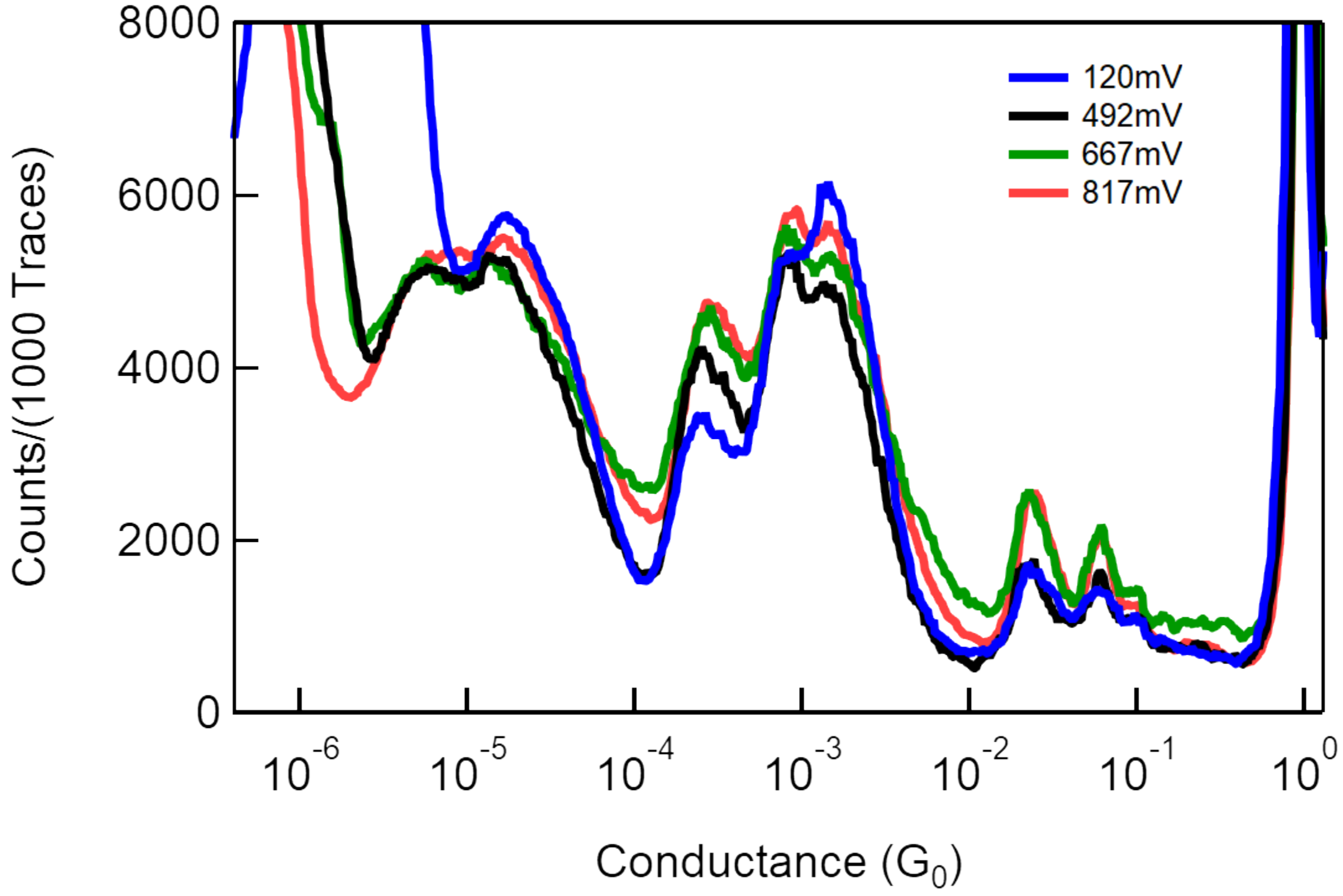

**Figure S2.** Conductance histograms of [KAu(CN)2] from measurements performed at tip biases of 120 mV (blue), 492 mV (black), 667 mV (green), and 817 mV (red).

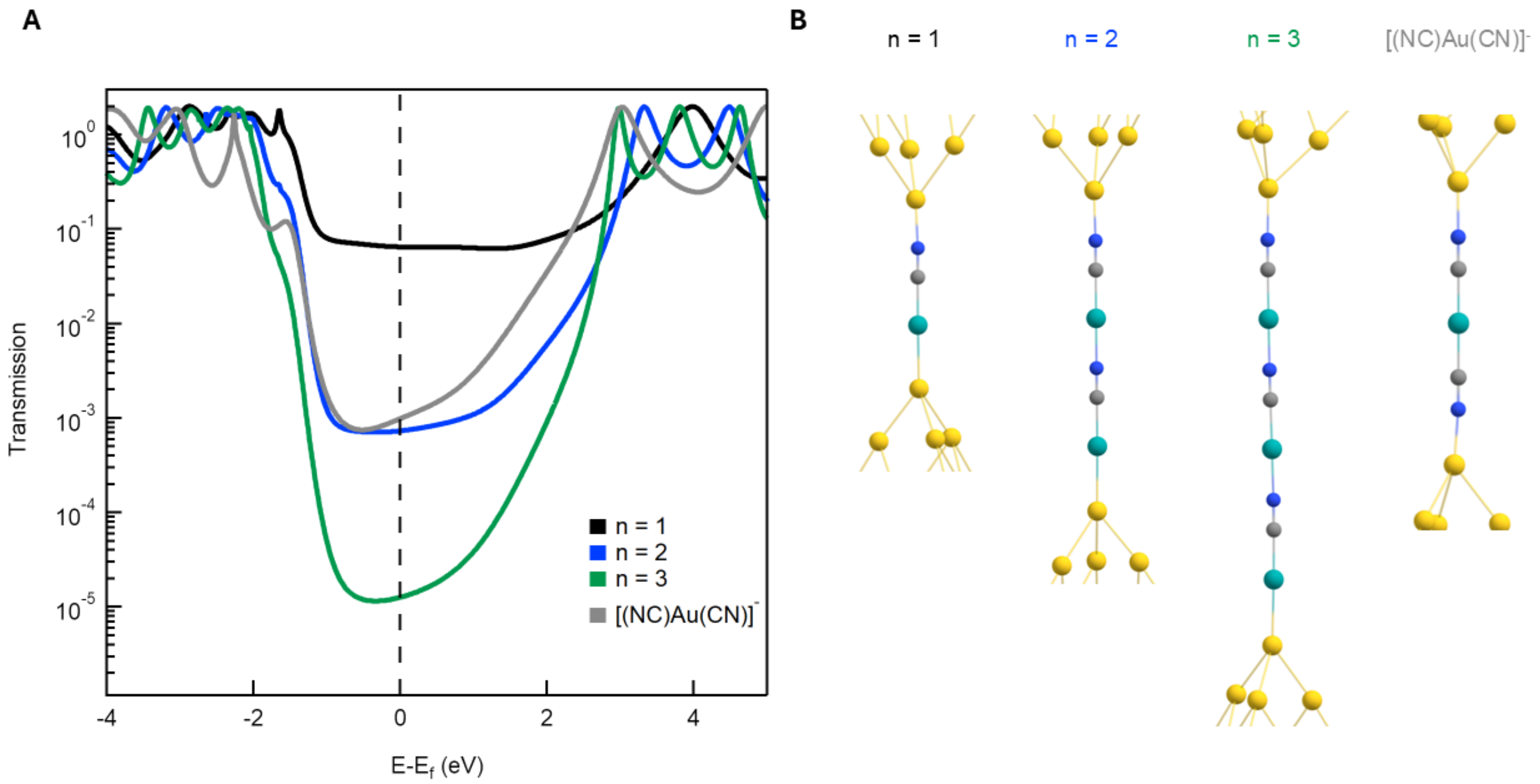

**Figure S3.** ) Transmission curves for $(Au(CN))_n$ *n* = 1-3 and $[(NC)Au(CN)]^-$ C) DFT relaxed structures for $(Au(CN))_n$ n = 1-3 and $[(NC)Au(CN)]^-$ with $Au_{34}$ electrodes.

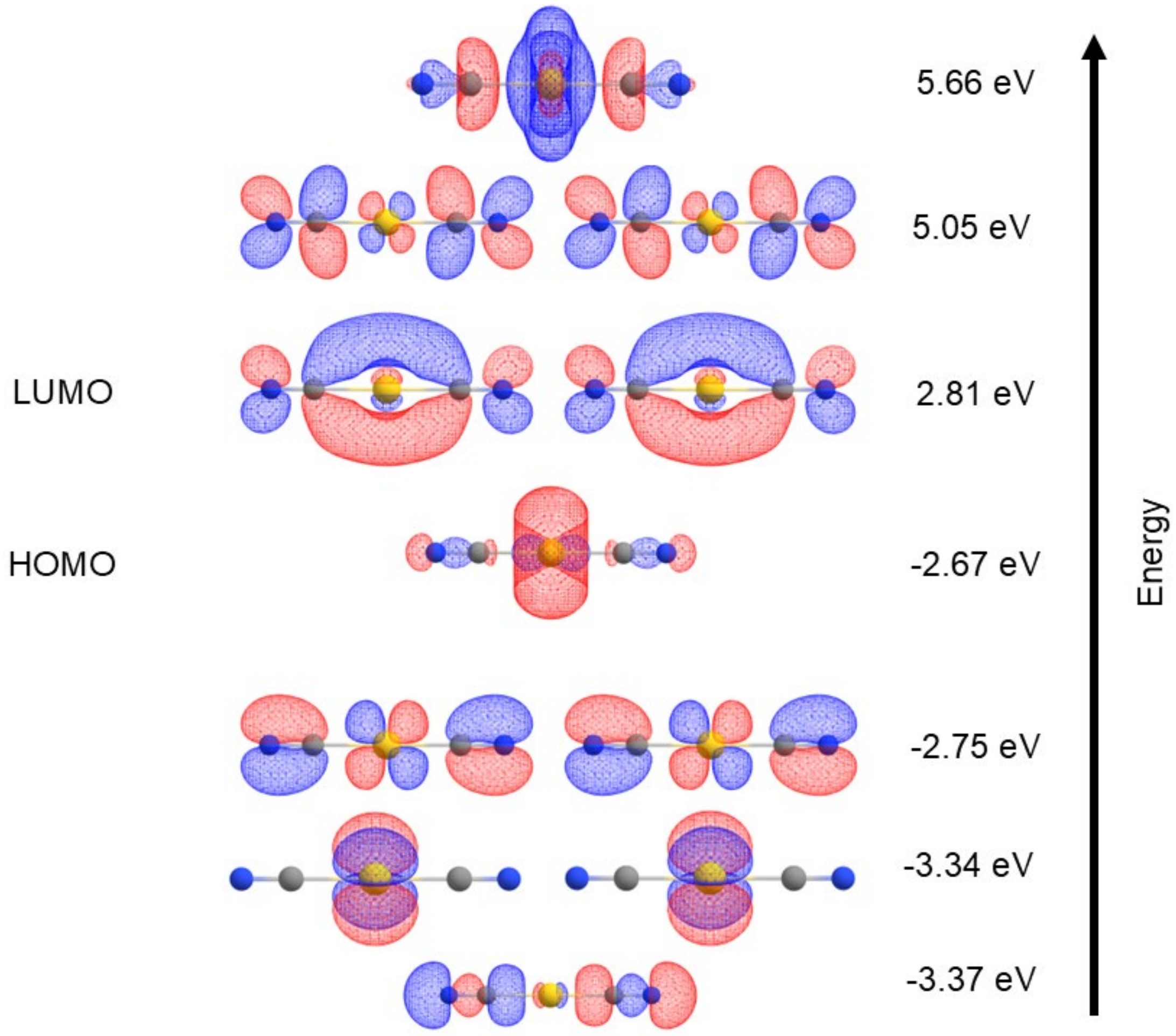

**Figure S4.** Isosurface plots of $[(NC)Au(CN)]^-$ gas phase molecular orbital at a contour level of 0.1.

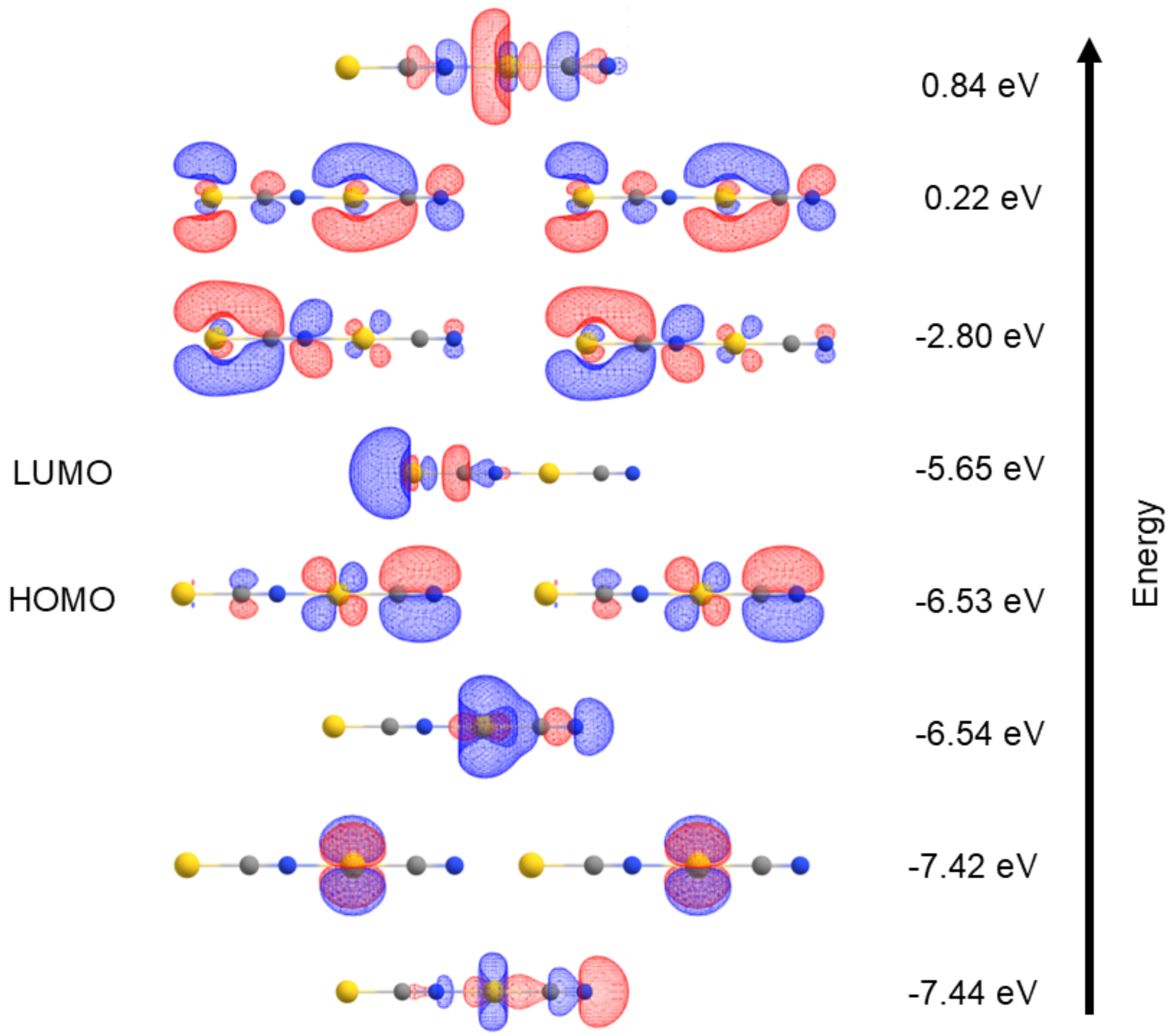

**Figure S5.** Isosurface plots of [Au(CN)Au(CN)]  gas phase molecular orbital at a contour level of 0.1.

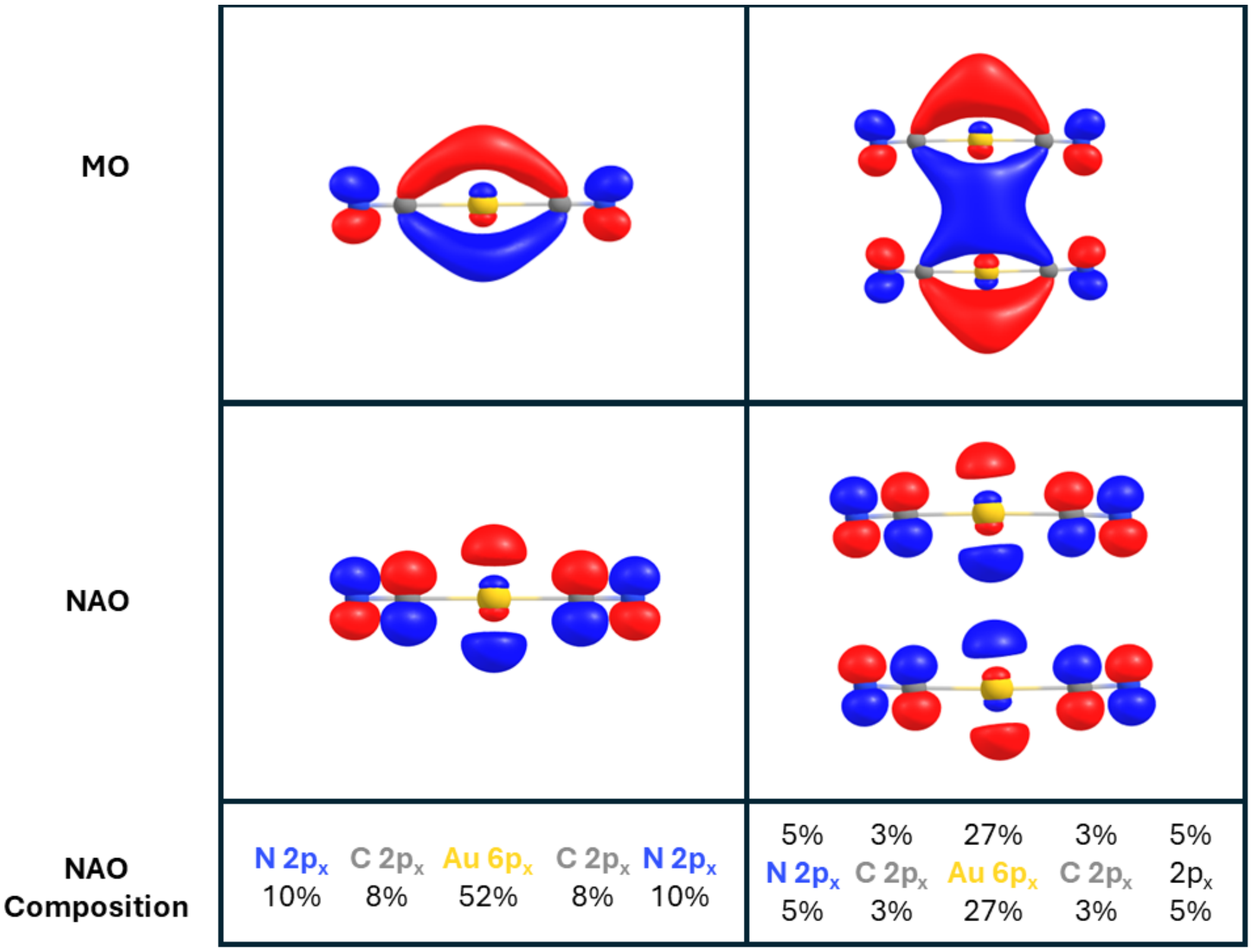

**Figure S6.** (Top) Isosurface of $\{Au(CN)_2\}^-$ wires from (Left, monomer) E2 and (Right, dimer) E4 eigenstates, calculated in Gaussian [ref] with the PBEPBE functional and def2-QZVPD basis set. (Middle) Isosurface plots of select natural atomic orbitals (NAO) for monomer (left) and dimer (right). (Bottom, Left) NAO composition of monomer. (Bottom, Right) NAO composition of dimer structure with top/bottom monomer composition listed above/below, respectively.

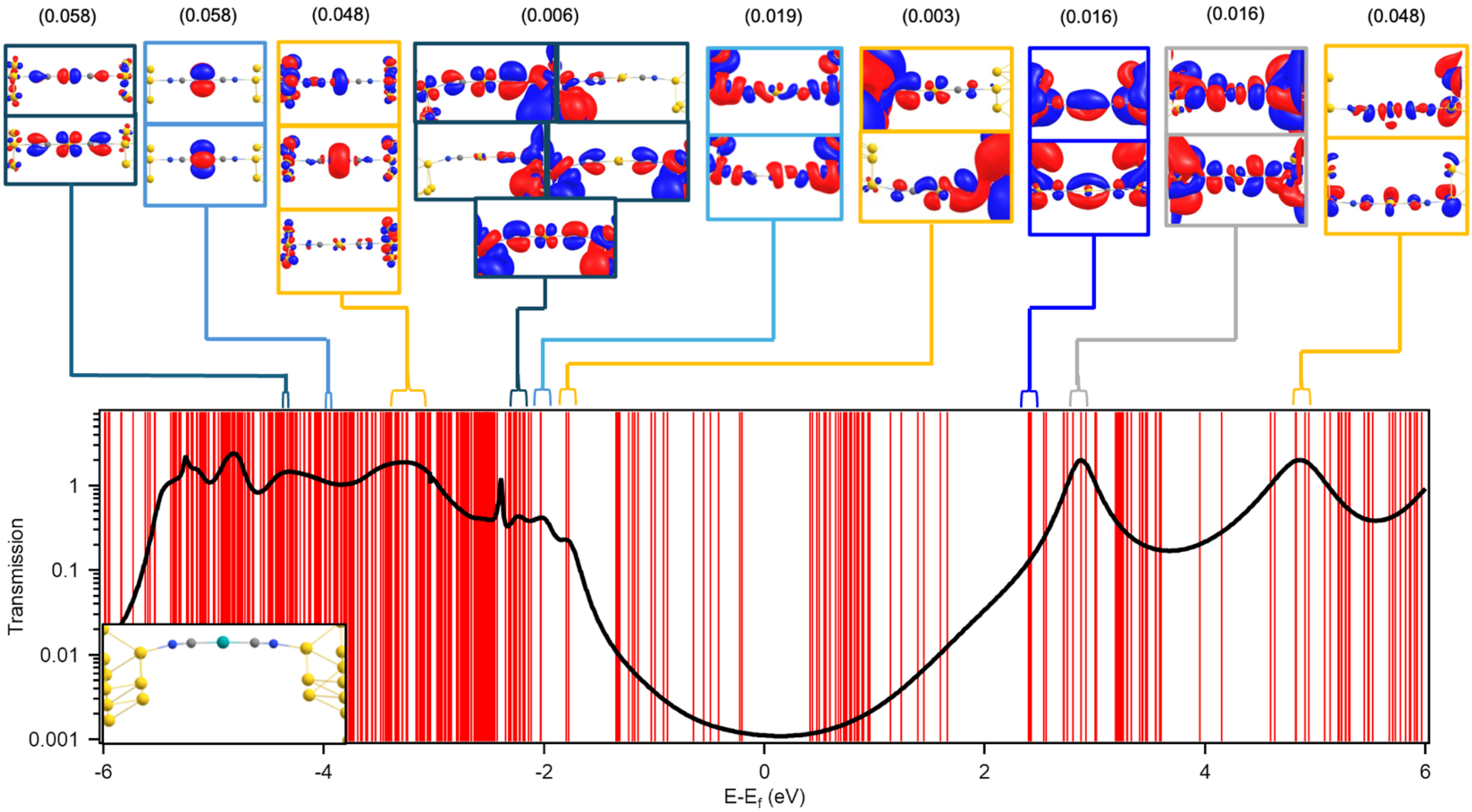

**Figure S7**. Eigenstate Analysis of Au32-{(NC)Au(CN)}⁻-Au32 Junction (Top) Selected isosurface plots of eigenstates with labeled contour level above the eigenstates. (Bottom) Transmission spectrum with energies of all eigenstates labeled with red vertical lines.

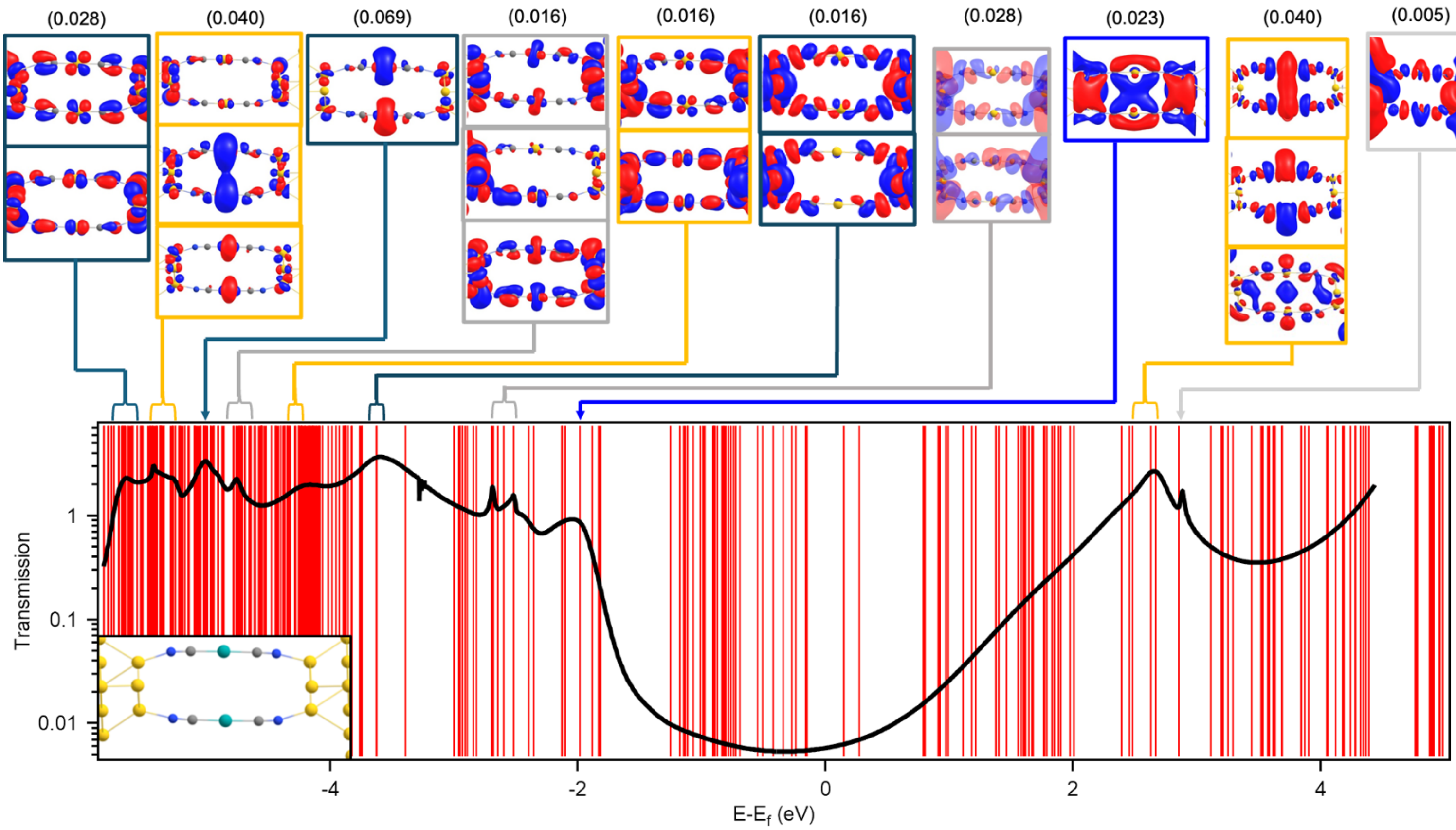

**Figure S8**. Eigenstate Analysis of $Au_{32}$-2{(NC)Au(CN)}$^-$-$Au_{32}$ Junction (Top) Selected isosurface plots of eigenstates with labeled contour level above the eigenstates. (Bottom) Transmission spectrum with energies of all eigenstates labeled with red vertical lines. We note that the calculated transmission of the dimer junction shown here and in Figure 3B is $5.80 \times 10^{-3}$, which is 5.18 times the monomer conductance in Figure 3B and S7 and significantly overestimates the measured M2/M1 ratio of 2.7, as listed in Table 2. Furthermore, the symmetric geometries of (NC)Au(CN) in Figure 3B cannot explain the conductance enhancement in the HG region, which contains a single Au(CN) unit and thus lacks a central Au atom for forming 6pσ MOs. We conclude that although the geometries in Figure 3 on sharp electrodes qualitatively capture the conductance enhancement due to aurophilicity, they are unlikely to account for all the measured conductance signatures in H, M and L regions.

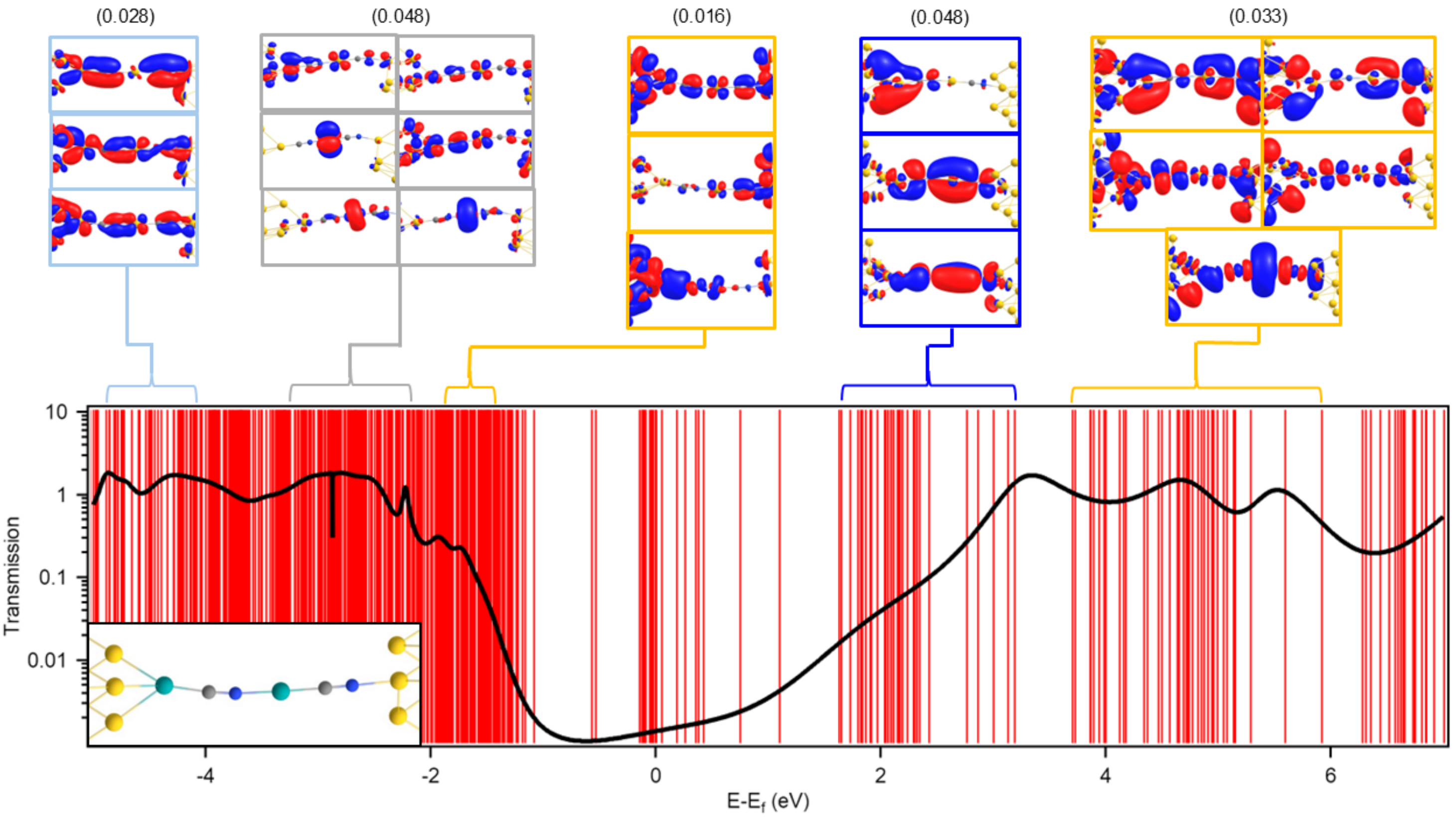

**Figure S9**. Eigenstate Analysis of $Au_{32}$-$(AuCN)_2$-$Au_{32}$ Junction (Top) Selected isosurface plots of eigenstates with labeled contour level above the eigenstates. (Bottom) Transmission spectrum with energies of all eigenstates labeled with red vertical lines.

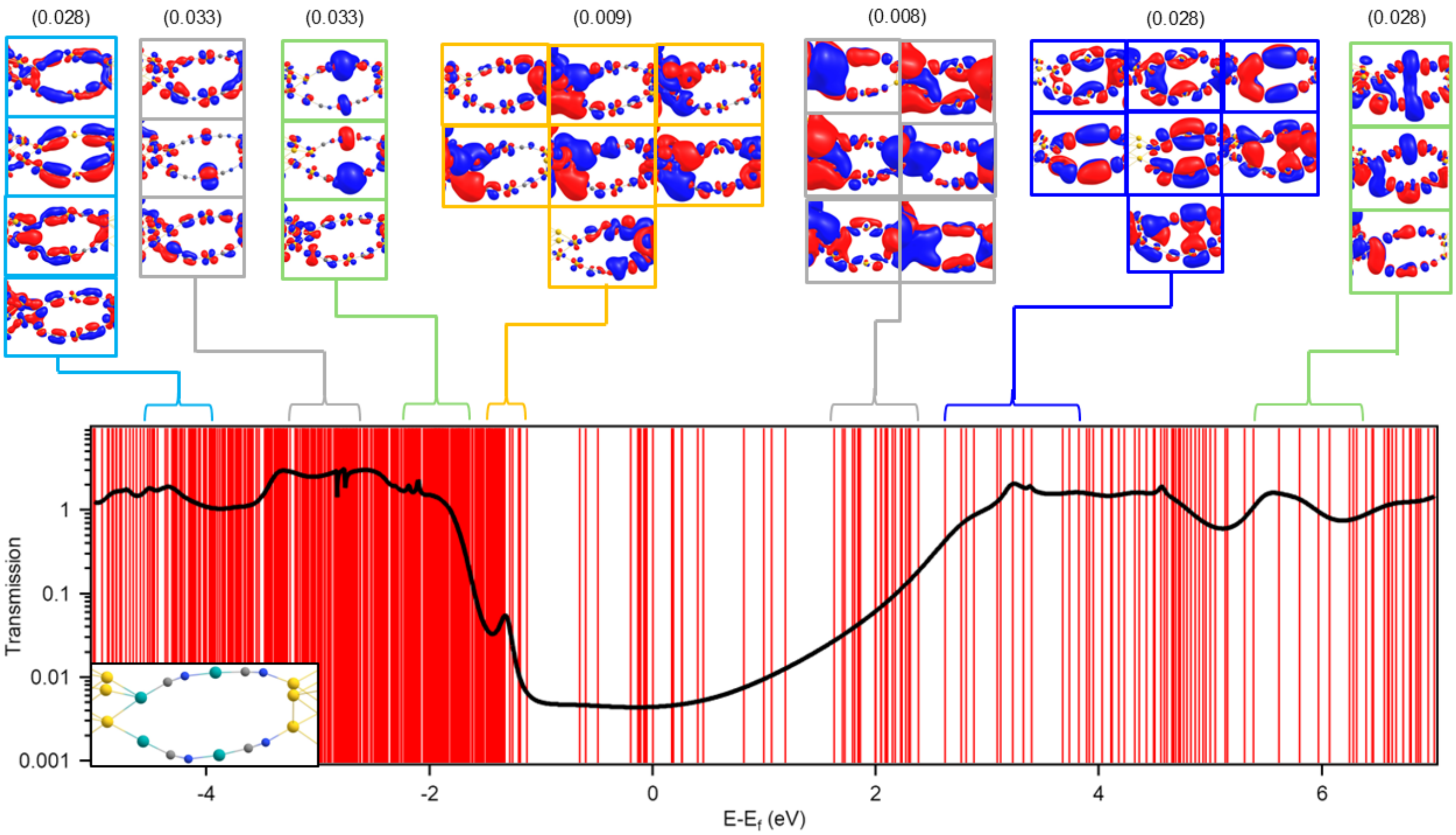

**Figure S10**. Eigenstate Analysis of $Au_{32}$-2$(AuCN)_2$-$Au_{32}$ Junction (Top) Selected isosurface plots of eigenstates with labeled contour level above the eigenstates. (Bottom) Transmission spectrum with energies of all eigenstates labeled with red vertical lines. The decrease in aurophility between gold atoms in the middle of the complex is attributed to greater internuclear distance between the centers and the development of a torsion angle between the overlapping 6p orbitals.

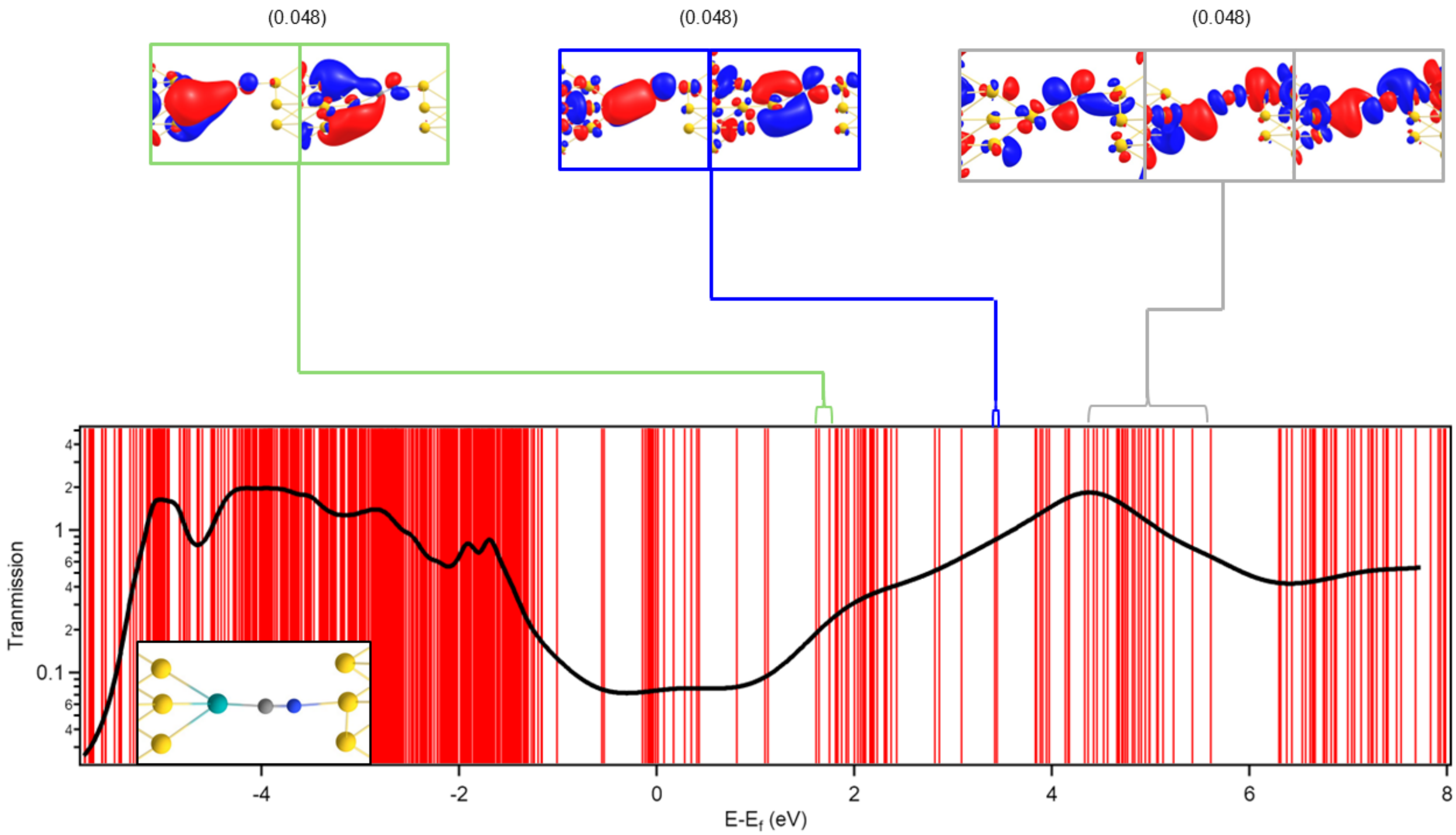

**Figure S11.** Eigenstate Analysis of $Au_{32}$-(AuCN)-$Au_{32}$ Junction (Top) Selected isosurface plots of eigenstates with labeled contour level above the eigenstates. (Bottom) Transmission spectrum with energies of all eigenstates labeled with red vertical lines.

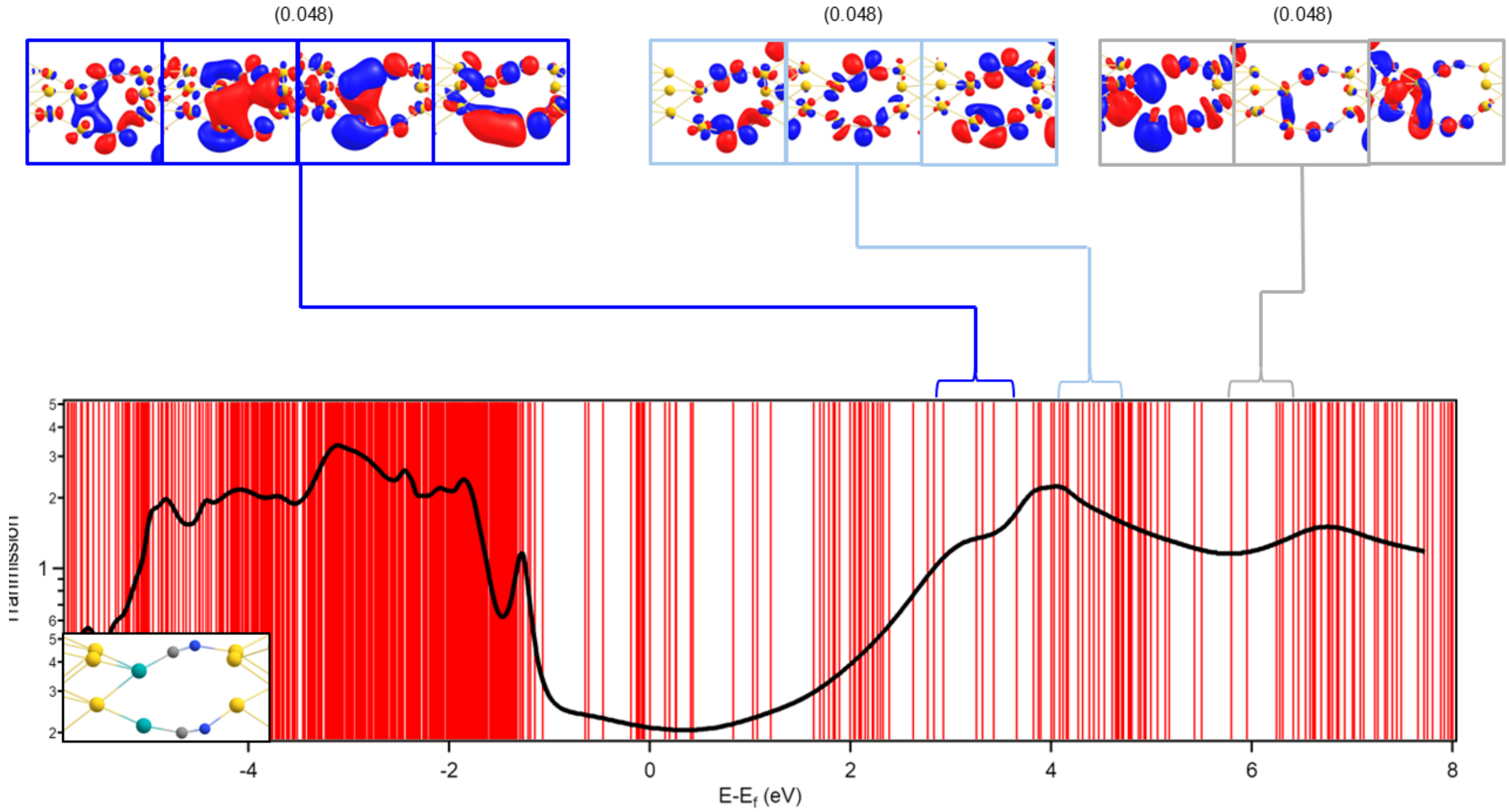

**Figure S12.** Eigenstate Analysis of $Au_{32}$-2(AuCN)-$Au_{32}$ Junction (Top) Selected isosurface plots of eigenstates with labeled contour level above the eigenstates. (Bottom) Transmission spectrum with energies of all eigenstates labeled with red vertical lines.

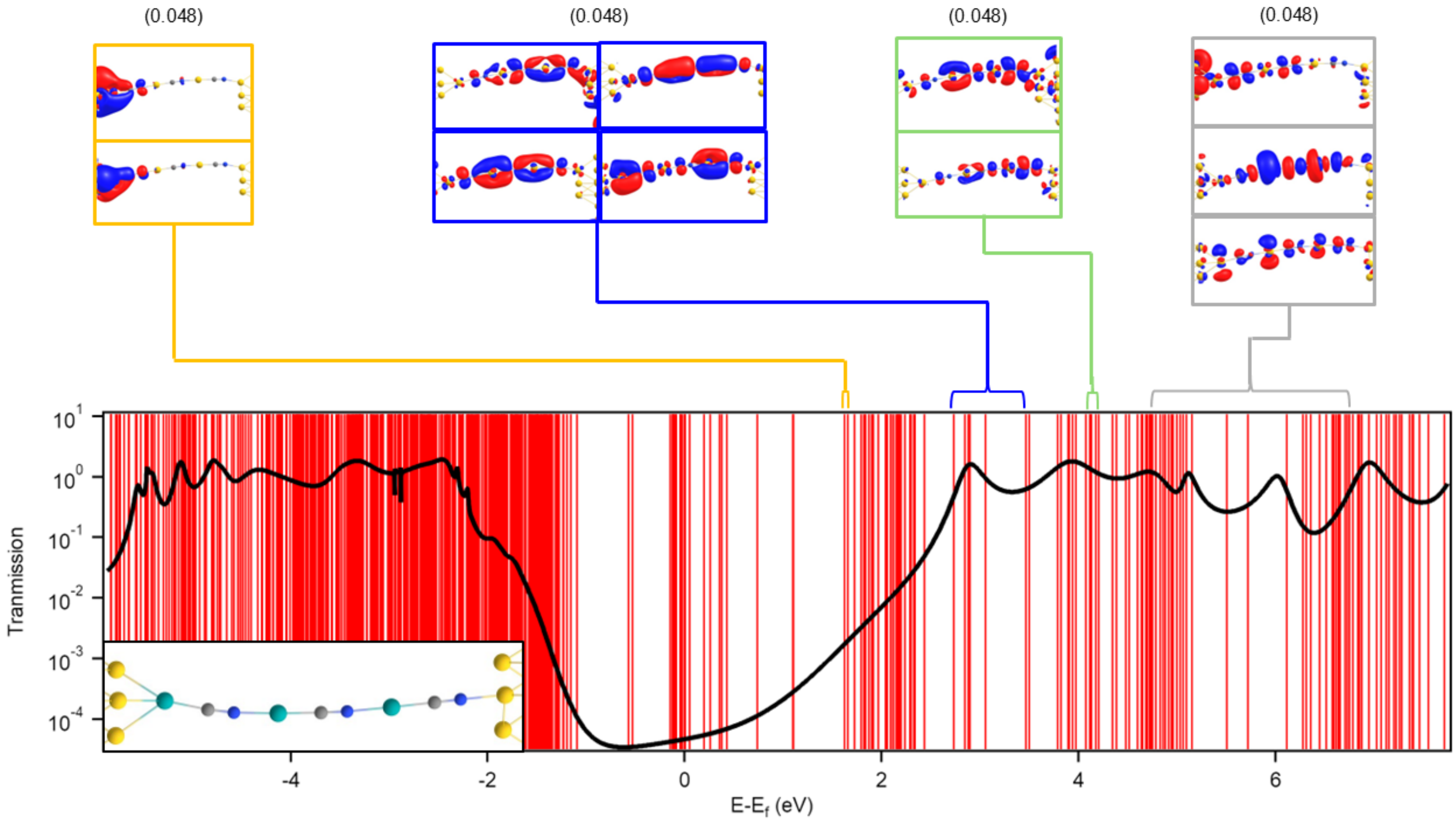

**Figure S13.** Eigenstate Analysis of $Au_{32}$-$(AuCN)_3$-$Au_{32}$ Junction (Top) Selected isosurface plots of eigenstates with labeled contour level above the eigenstates. (Bottom) Transmission spectrum with energies of all eigenstates labeled with red vertical lines.

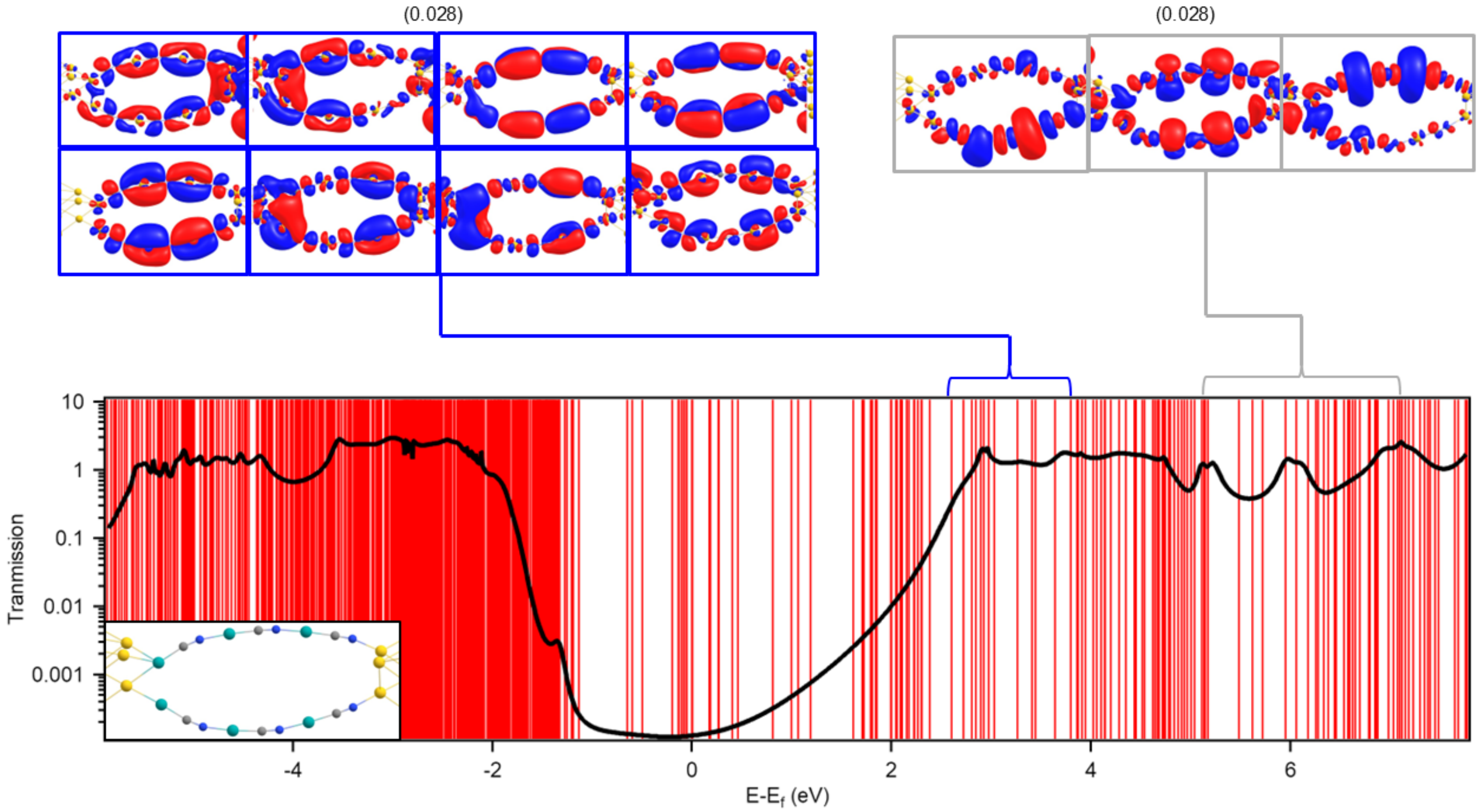

**Figure S14.** Eigenstate Analysis of $Au_{32}$-2$(AuCN)_3$-$Au_{32}$ Junction (Top) Selected isosurface plots of eigenstates with labeled contour level above the eigenstates. (Bottom) Transmission spectrum with energies of all eigenstates labeled with red vertical lines.

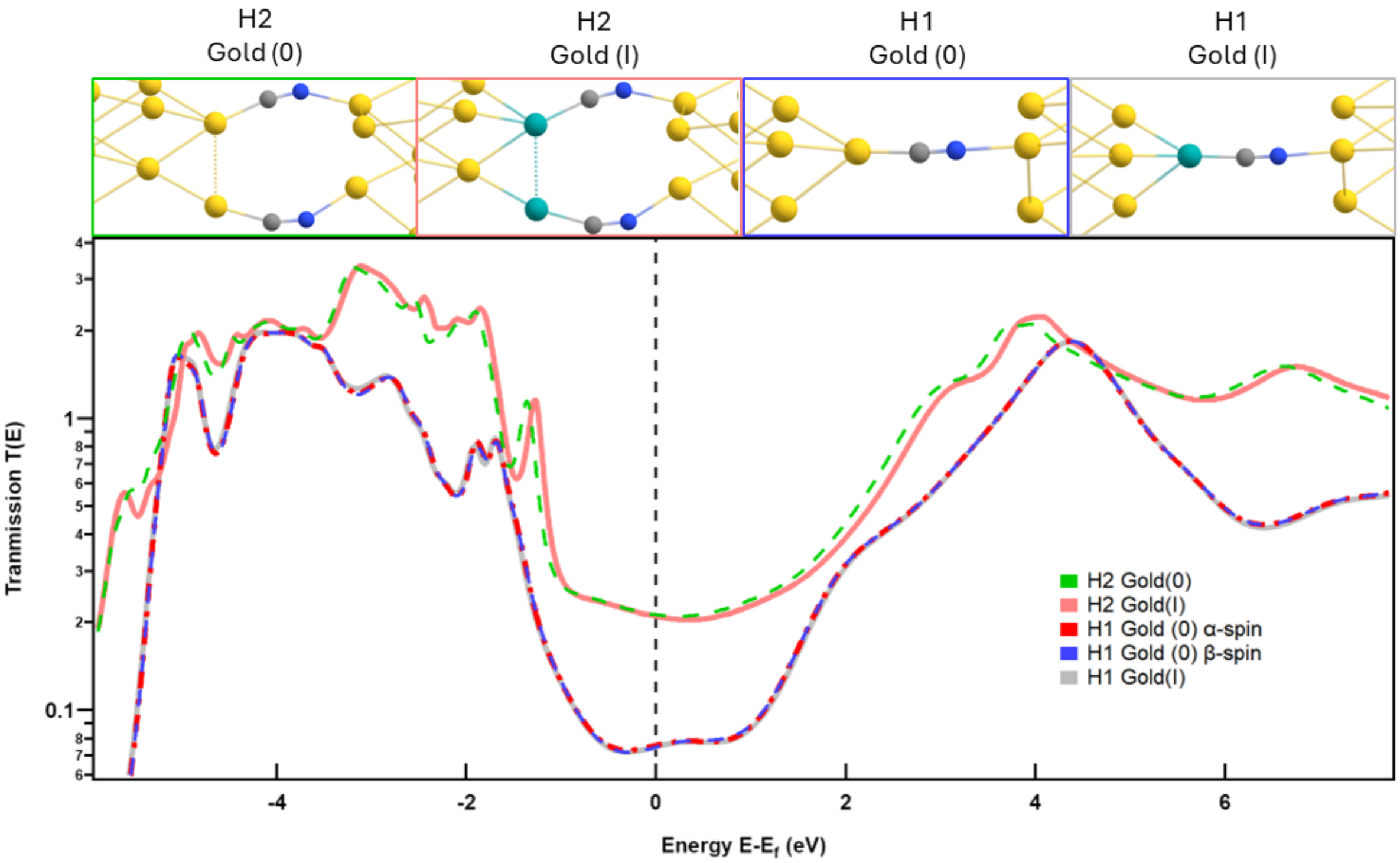

**Figure S15.** (Top) DFT relaxed structures for 2{Au(CN)}$^{2-}$ (H2 Gold(0)), 2{Au(CN)} (H2 Gold(I)), {Au(CN)}$^{-}$ (H1 Gold (0)), and {Au(CN)} (H1 Gold (I))) with $Au_{32}$ electrodes. (Bottom) Corresponding transmission spectra for these structures.